\documentclass[prl,twocolumn,showpacs]{revtex4-1}
\usepackage{graphicx}
\usepackage{color}
\usepackage{amssymb,amsmath}
\usepackage{bm}
\usepackage[applemac]{inputenc}
  
\newcommand{\sect}[1]{\textit{#1.---}\ignorespaces} 
 
\begin{document}     
  
\title{Helical networks in twisted bilayer graphene under interlayer bias}
\author{Pablo San-Jose$^1$, Elsa Prada$^2$}
\affiliation{$^1$Instituto de Ciencia de Materiales de Madrid (ICMM-CSIC), Cantoblanco, 28049 Madrid, Spain\\
$^2$Departamento de Física de la Materia Condensada, Instituto de Ciencia de Materiales Nicolás Cabrera, Instituto de Física de la Materia Condensada, Universidad Autónoma de Madrid, Cantoblanco, 28049 Madrid, Spain}

\date{\today} 

\begin{abstract} 
A twisted graphene bilayer exhibits a triangular Moiré pattern in the local stacking, that smoothly alternates between the three basic types AA', AB' and BA'. Under an interlayer bias $U$, the latter two types develop a spectral gap, characterised by opposite valley Chern numbers. We show that for large enough Moiré periods and bias, 
these regions become depleted electronically, and topologically protected helical modes appear at their boundaries. This gives rise to a delocalised topological network of the Chalker-Coddington type, composed of valley current vortices. This network can be tailored by controlled deposition of valley-mixing adsorbates, which block transmission in selected links, thus opening the possibility of custom topological nanoelectronics.
 
\end{abstract}

\maketitle
 
Topological phases of matter have become the center of a very active research field in condensed matter physics \cite{Hasan:RMP10, Qi:RMP11, Bernevig:13}. One of the main appeals of these systems is the appearance, due to robust topological reasons, of helical or chiral \footnote{The terms ``helical'' and ``chiral'' are used for the cases with and without time reversal invariance, respectively.}  metallic states on the boundary of an otherwise insulating material. These states are protected against backscattering under very general conditions, and could be used to transport charge, spin, or information in general, in a directed and dissipationless fashion between distant regions. In other words, topological edge channels are the ultimate interconnects of nanoelectronic systems. While this idea has been already extensively studied \cite{Qi:PT10}, here we propose a step up in complexity towards the possibility of designing and controlling \emph{networks} of topological channels. Since channels are conventionally pinned to the boundaries of the system, the problem of scaling them up into a complete topological circuit is far from trivial. However, and quite remarkably, nature can provide us with such a possibility in a ``built-in" manner: by applying an opposite electric bias to each layer in a twisted graphene bilayer, we find that a network of helical electronic states develops along stacking boundaries (solitons) inherent to the system.  A direct imaging of stacking solitons in such bilayers has recently been reported experimentally by Alden at al. \cite{Alden:PNAS13}. We show, furthermore, that the network of helical channels studied here can can be ``programmed" into a broad range of specific circuits by the controlled deposition of valley-mixing adsorbates that shut off selected network links.

It is known that an interlayer bias applied to a Bernal-type stacking of two graphene layers (also known as AB'or BA' stacking \footnote{The notation for minimal graphene stackings refers to the four carbon atoms A, B, A' and B' of the minimal unit cell of the bilayer. Thus, AB' stacking refers to the case in which the A carbon in the lower layer is vertically aligned with the B atom in the upper layer, whilst AA' has both A and B atoms of each layer vertically aligned. Clearly, non-minimal stackings are also possible, such as that of a generic twisted bilayer.}), opens a band gap around each valley\cite{McCann:PRB06,Castro:JPCM10}. Gaps up to $250$ meV have been achieved experimentally \cite{Mak:PRL09,Zhang:N09}. 
Beyond its potential application in nanoelectronics, the gap tunability of graphene bilayers is also intriguing from a fundamental point of view. Firstly, it depends critically on the precise AB' or BA' interlayer stacking. An AA' bilayer, for example, is a good metal for any reasonable interlayer bias. Secondly, the gap can be made to ``change sign" by inverting the bias $U$ into $-U$. Performing such an inversion between two adjacent regions in space gives rise to two topologically protected helical (TPH) modes per valley and spin, that flow without resistance along the interface between the two regions \cite{Martin:PRL08, Nunez:APL11, Qiao:NL11} \footnote{The helical property of protected modes in the context of graphene bilayers is the fixed relation between their direction of propagation along the boundary and their valley quantum number.}. Analogous topologically protected channels arise in a range of systems, in which the gapped local band structure changes topology across a boundary, e.g. at the edges of Hall bars, or the surface of topological insulators \cite{Qi:PRB08,Hasan:RMP10}. 
Lastly, and rather surprisingly, a change of the topology of a tuned gap can be induced also with a \emph{uniform} bias $U$ by smoothly transitioning from AB' to BA' stackings \cite{Wright:APL11, Vaezi:PRX13, Zhang:PNAS13}, which are related by mirror symmetry. This can be achieved, e.g., by applying a different strain to each layer \cite{Vaezi:PRX13}, and also gives rise to TPH modes. 

In this work we study the formation of TPH modes in graphene bilayers with a generic (non-minimal) stacking. These are known as twisted graphene bilayers, and are characterised by a finite interlayer rotation angle. We find that TPH modes may arise in these systems \emph{without} strain under a \emph{uniform} interlayer bias $U$. These helical modes arrange into a triangular network of valley-polarized current running along zero mass lines \cite{Tudorovskiy:PRB12}. This is a variation of what is commonly known as a Chalker-Coddington network \cite{Chalker:JPCSSP88,Ho:PRB96,Mkhitaryan:PRB09}. Superimposed onto this delocalized network, a discrete set of states strongly localized away from AB'/BA' regions emerge at resonant energies. Such restructuring of electronic states occurs for energies $|\epsilon|<|U|/2$, and produces clearly measurable signatures in the density of states (DOS) and the optical conductivity $\sigma_{xx}(\omega)$ \cite{Stauber:PRB08a,Nair:S08} (see Appendix for details on the latter). Interestingly, the mechanism that gives rise to the helical network also appears, although for different reasons, in graphene monolayers on boron-nitride (h-BN) substrates \cite{Sachs:PRB11,Ponomarenko:12,Tudorovskiy:PRB12,Mucha-Kruczynski:13,Kindermann:PRB12}. The lattice mismatch between graphene and h-BN, however, does not allow for an arbitrarily large Moiré period, so that unlike in twisted bilayers, the helical network is not expected to fully develop in that case.
\begin{figure}[t] 
\centering
\includegraphics[width=\columnwidth]{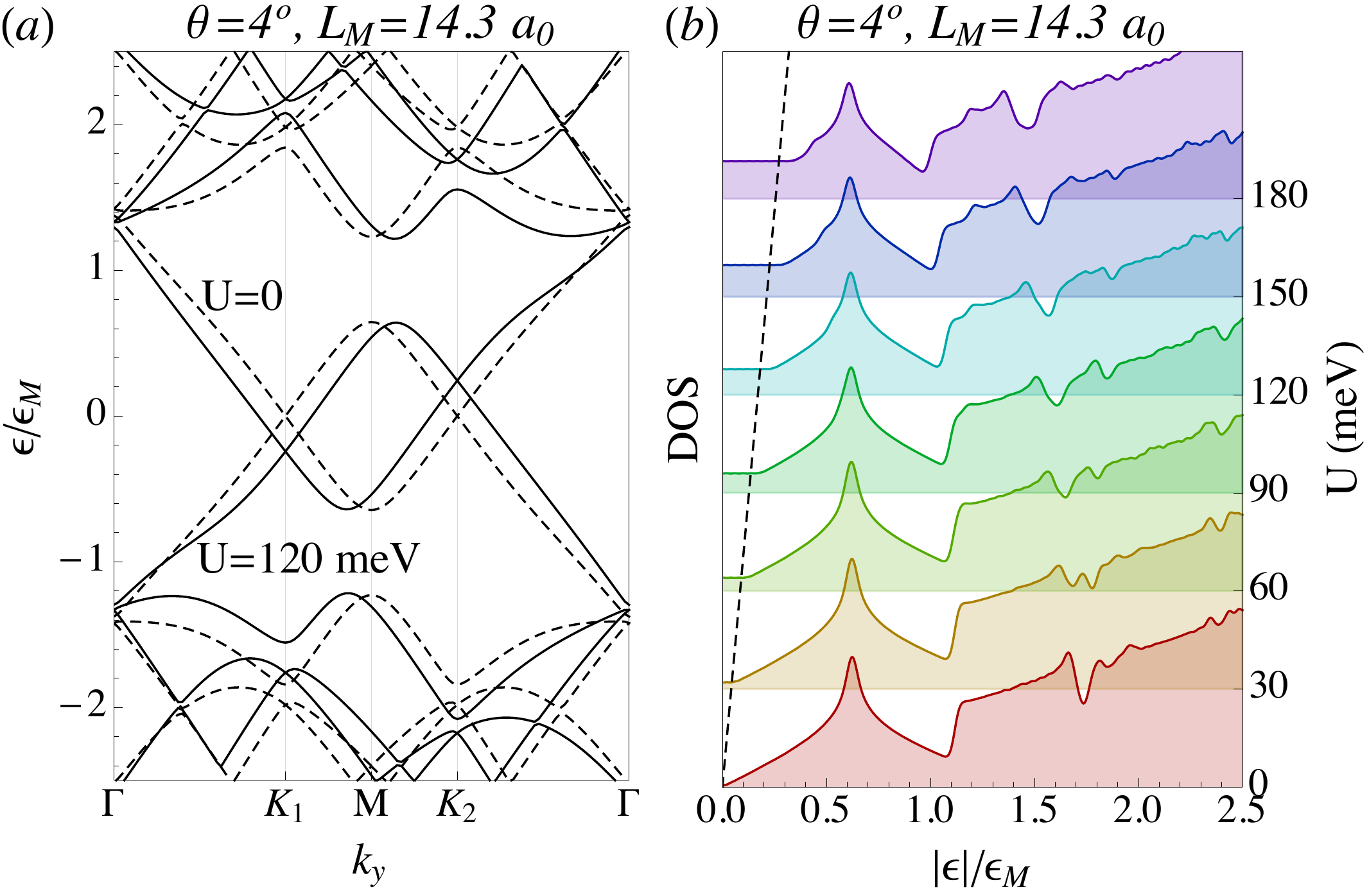} 
\caption{(Color online) (a) Bandstructure and (b) density of states (DOS) for a $\theta=4º$ twisted bilayer with and without interlayer bias $U$. At such angles, a bias merely shifts the two Dirac cones, creating a flat DOS plateau for $|\epsilon|<|U|/2$ [to the left of the dashed line in (b)].} \label{fig:large}
\end{figure}

\sect{Band topology and helical modes}
Let us briefly review some concepts on the band topology of graphene bilayers. Any electronic band defined in a finite Brillouin zone (BZ) has an associated integer topological invariant \cite{Avron:PRL83,Zak:PRL89,Hasan:RMP10,Xiao:RMP10}, known as Chern number $C$, which determines e.g. its contribution to the Hall conductivity $\sigma_{xy}=Ce^2/h$ when the band is completely filled \cite{Laughlin:PRB81,Thouless:PRL82,Kohmoto:AOP85,Kohmoto:PRB89,Avron:PT03}. In 2+1 dimensions, the Chern number is defined in terms of eigenstates $|\psi(\bm{k})\rangle$ as $C=\frac{1}{2\pi}\int_{BZ} d^2k \,F(\bm{k})$, where $F(\bm{k})=2\mathrm{Im}\langle \partial_{k_x}\psi(\bm{k})|\partial_{k_y}\psi(\bm{k})\rangle$ is the Berry curvature. A biased AB'-stacked graphene bilayer develops a gap around its two valleys, located at points $K$ and $K'$ in the BZ. The Berry curvature of the valence band is peaked around these two points and has opposite sign, so that $C=0$. However, a (non-quantized) valley Chern number may be defined by confining the $\bm{k}$ integral to the region around a single valley, $C_{K^{(')}}=\frac{1}{2\pi}\int_{K^{(')}} d^2k \,F(\bm{k})$. If the bias is small as compared to the interlayer coupling, $U<\gamma_\perp$, $C_{K}=-C_{K'}\approx \textrm{sign}(U)$ is approximately quantised. This valley Chern number plays then the role of a proper topological charge \cite{Prada:SSC11, Zhang:PNAS13}, as long as valleys are kept decoupled by any perturbation (i.e. as long as the bilayer, and $U(\bm{r})$ itself, are smooth on the scale of the lattice constant $a_0$). In particular, it implies the emergence of two TPH modes per valley and spin along an interface where $U$ (and hence $C_{K^{(')}}$) smoothly changes sign \cite{Martin:PRL08}. Likewise, since a BA'-stacking has opposite valley Chern numbers than AB' ($C^{AB'}_{K^{(')}}=-C^{BA'}_{K^{(')}}$), a smooth transition between these two stackings will also produce two TPH modes per valley and spin \cite{Wright:APL11}. This result is a version of the bulk-boundary correspondence, which states that $N=|C_2-C_1|\,\mathrm{mod}\,2$ TPH modes emerge where the Chern number changes from $C_1$ to $C_2$. The mod 2 is absent in our case ($N=2$) since we consider only valley-preserving perturbations, hence ``weak'' topological protection, see Ref. \onlinecite{Hasan:RMP10}. A transition between stackings arises naturally in a bilayer with a relative rotation between layers, the so-called twisted graphene bilayer.

\begin{figure*}[t] 
\centering
\includegraphics[width=\textwidth]{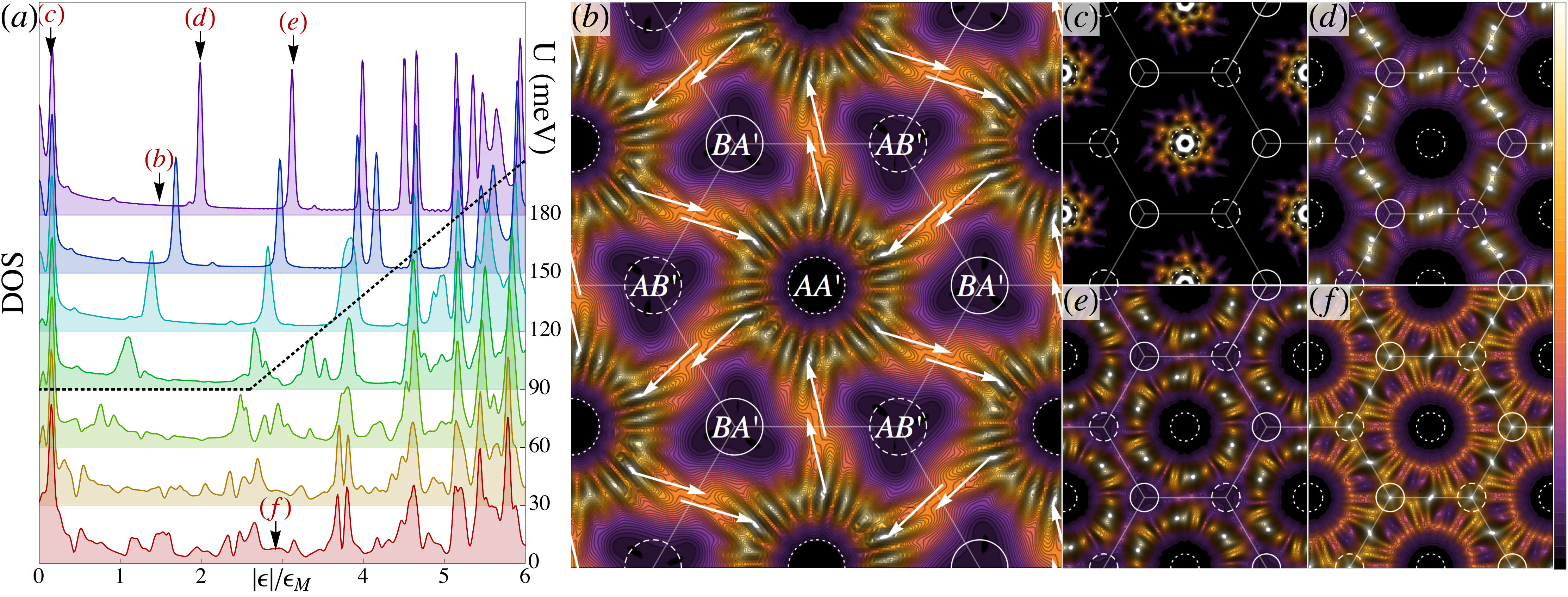}
\caption{(Color online)  (a) Density of states of a $L_M=290 a_0$ ($\theta\approx 0.2º$) twisted bilayer with increasing bias $U$. The region above the dotted line denotes the helical network regime. (b-f) Periodic spatial profile of electron probability density for several characteristic eigenstates in the spectrum [arrows in (a)], all chosen at one of the $K$ points. Black colour corresponds to zero density, and white to maximum. Pannel (b) shows a fully developed helical network (Chalker-Coddington) state, with arrows indicating the orientation of the valley currents. Pannels (c), (d) and (e) show various resonant states, all localized away from AB'/BA' regions by the finite bias. Pannel (f) shows a $U=0$ delocalized state.} \label{fig:eigenstates}
\end{figure*}

\sect{Continuum theory of twisted bilayers}
A twisted graphene bilayer is parameterised by the relative rotation angle $\theta$, with $\theta=0$ denoting the AA' bilayer. Geometrically, such rotation produces a periodic triangular Moiré pattern in the local stacking, which smoothly alternates between the three minimal types AA', AB' and BA'. For an angle of rotation $\theta<30º$, the Moiré pattern has a period $L_M=a_0/\left[2\sin(\theta/2)\right]$ ($a_0=2.46 \textrm{\AA}$ is the monolayer Bravais period). Electronically, the low energy sector of the system does not depend on the detailed crystallography \cite{Santos:PRL07,Mele:PRB10,Mele:PRB11,Gail:PRB11,Santos:PRB12}, and is well described by a continuum theory in which the two valleys are decoupled, and the Dirac points in each layer are offset by a momentum $\Delta K=2|K|\sin(\theta/2)$, with $|K|=4\pi/3a_0$. In the absence of interlayer coupling, the two cones intersect around the M point of the superlattice, at energy $\epsilon_M(\theta)=\hbar v_F\Delta K/2$ (which is $\approx 88\mathrm{meV} \times\theta$(deg) for small angles). This is the natural energy scale in the low energy sector. The coupled bilayer is described within each valley by the continuum spin-degenerate Hamiltonian \[
H=\left(\begin{array}{cccc}
U/2 & \hbar v_F \Pi_+^\dagger & V^*(\bm{r}) & V^*(\bm{r}-\bm{\delta r}) \\
\hbar v_F\Pi_+ & U/2 & V^*(\bm{r}+\bm{\delta r}) & V^*(\bm{r})    \\
V(\bm{r}) & V(\bm{r}+\bm{\delta r}) & -U/2 & \hbar v_F \Pi_-^\dagger\\
V(\bm{r}-\bm{\delta r}) & V(\bm{r}) & \hbar v_F\Pi_- & -U/2   \\
\end{array}\right)
\]
which is written in the $A, B, A', B'$ basis. Here $\Pi_\pm=k_x+i(k_y\mp\Delta K/2)$, and the function $V(\bm{r})=\frac{1}{3}\gamma_\perp\sum_{i=1}^3 e^{i\bm{g}_i\cdot\bm{r}}$ describes the periodic spatial variation of the interlayer coupling, with $\gamma_\perp\approx 0.33 \mathrm{eV}$. The Moiré lattice vectors are denoted by $\bm{a}_{1,2}$ (so $L_M=|\bm{a}_{1,2}|$), and conjugate momenta are $\bm{g}_{1,2}$, such that $\bm{g}_i\cdot\bm{a}_j=2\pi \delta_{ij}$ (we also define $\bm{g}_3=0$). The AA' sublattice is centered at the origin, with the AB'/BA' sublattices offset by $\pm\bm{\delta r}=(\bm{a}_1-\bm{a}_2)/3$.

The continuum model has two distinct regimes, depending on the value of dimensionless parameter $\alpha(\theta)=\frac{1}{6}\gamma_\perp/\epsilon_M(\theta)$. The regime of large angles ($\theta\gg 1º$, $\alpha\ll 1$) is amenable to perturbation theory in $\alpha$, and has been extensively studied. It is characterised by the formation of a van-Hove singularity \cite{Santos:PRL07,Li:NP10,Luican:PRL11,Yan:PRL12, Brihuega:PRL12,Moon:PRB13,Tabert:PRB13,Zou:PRL13} at energy $\epsilon_\textrm{vH}\approx\epsilon_M(1-2\alpha)$ close to the M point [see peak in the DOS at $\sim 0.6 \epsilon_M$ in Fig. \ref{fig:large}(b)] and a suppression of the Fermi velocity $v_F^*\approx v_F(1-9\alpha^2)$ \cite{Santos:PRL07,Luican:PRL11}. 
In the presence of a finite bias $U$, the band structure in the large angle regime reacts in a predictable fashion, merely shifting the two Dirac cones by $\pm U/2$, see Fig. \ref{fig:large}(a), since their eigenstates mostly reside in different layers in this case. This results in a low energy density of states that is flat instead of linear for $|\epsilon|<|U|/2$, see Fig. \ref{fig:large}(b). The DOS profile is reminiscent of the perfect AA'-stacked bilayer's, whose band structure is also composed of two shifted Dirac cones. Above $|U|/2$, the DOS is largely unaffected by the bias.

The regime of small angles has been less explored and has a much more complex structure. In the unbiased case (see Appendix), secondary van Hove singularities above the first appear and move lower in energy as $\theta$ is decreased. As each singularity approaches zero energy, it morphs into a quasi-flat miniband, 
which is accompanied by a vanishing Fermi velocity. This occurs at specific (``magic") twist angles $\theta_i$, the highest being $\theta_1\approx 1.05º$ (which corresponds to $L_M\approx 52 a_0$) \cite{Bistritzer:P11,Laissardiere:PRB12}. In general however, this simple picture can only make sense of the spectrum for angles down to around $\theta\sim 0.35º$. Below this $\theta$, the $U=0$ electronic structure becomes rather intricate. With the notable exception of a resilient AA'-localized state \cite{Laissardiere:NL10,San-Jose:PRL12} pinned at around $\epsilon\sim 0.17\epsilon_M$ (which is evolved from the original large angle van-Hove singularity), the DOS profile, as shown in the bottom curve of Fig. \ref{fig:eigenstates}(a), exhibits no clear structure (although see Ref. \onlinecite{Laissardiere:PRB12}), and states are typically delocalised [Fig. \ref{fig:eigenstates}(f)].

The small-angle electronic structure becomes much simpler as $U$ is increased. The AB' and BA' regions develop a local gap, with opposite valley Chern number. If $U$ is above certain threshold, and these regions are large enough (small enough angles), the bias tends to electronically deplete them. As a consequence, strong confinement of states away from the depleted AB'/BA' regions becomes possible. A discrete set of sharp localized resonances appear in the spectrum, as seen in the top curve of Fig. \ref{fig:eigenstates}(a). These localized states arrange spatially in a variety of patterns [see Fig. \ref{fig:eigenstates}(c,d,e)], but always away from the AB'/BA' regions. For off-resonant energies, the DOS exhibits a weak and featureless background that is in marked contrast to the intricate small-angle, $U=0$ DOS profile, and to the Dirac-like linear background of the large-angle DOS. This weak background corresponds to delocalized TPH modes running along AB'/BA' interfaces, which develop due to the spatial modulation in the valley Chern number of the gapped regions. The collection of helical channels arrange in a triangular Chalker-Coddington helical network, as shown in Fig. \ref{fig:eigenstates}(b). Each channel carries opposite current for opposite valleys. The pattern of valley currents for a given sign of $U$ is fixed by the band topology, and is shown by white arrows.

This new electronic phase requires strong depletion in AB' and BA' regions to fully develop \cite{Tudorovskiy:PRB12}, which in turn requires large bias and small angles.
One can approximately quantify this condition by considering the simplified 1D problem of an abrupt AB'/BA' interface with a uniform bias. Using the two component  approximation (valid for $U<\gamma_\perp$), an interface mode decays like $\sim e^{-\lambda |x|}$, with $\lambda=\sqrt{U\gamma_\perp}/2v_F$ \cite{Martin:PRL08}. We look for the value of $U$ that results in an amplitude decay of the interface mode of at least, say, 95\%, within the radius of the Moiré BA' region of the twisted bilayer, $R=L_M/2\sqrt{3}$. This is half the distance between AB' and BA' region centers, see Fig. \ref{fig:eigenstates}(b). Inserting $x=R$ this results in the following condition 
for the development of the helical network
\begin{equation}
z^2\left(\frac{l_\perp}{L_M(\theta)}\right)^2<\frac{U}{\gamma_\perp}<1, \hspace{.5cm}|\epsilon|<\frac{|U|}{2}
\label{criterion}
\end{equation}
where $z\equiv -4\sqrt{3}\,\mathrm{ln}\,0.05=20.76$, and $l_\perp=\hbar v_F/\gamma_\perp\approx 7.3 a_0$ is the interlayer coupling length \cite{Snyman:PRB07}. 
 
This analysis suggests that for a realistic value of $U\sim 90$ meV, the network requires a Moiré period $L_M\gtrsim 290\,a_0$ ($\theta<0.2º$), the value chosen in Fig. \ref{fig:eigenstates}. For this angle, the range of $U$ and $\epsilon$ that satisfies the criterion has been marked by a dotted boundary in Fig. \ref{fig:eigenstates}(a). We see the boundary roughly correlates with the crossover into the helical network regime, characterised by sharp DOS peaks on a weak smooth background.

\begin{figure}[t] 
\centering
\includegraphics[width=\columnwidth]{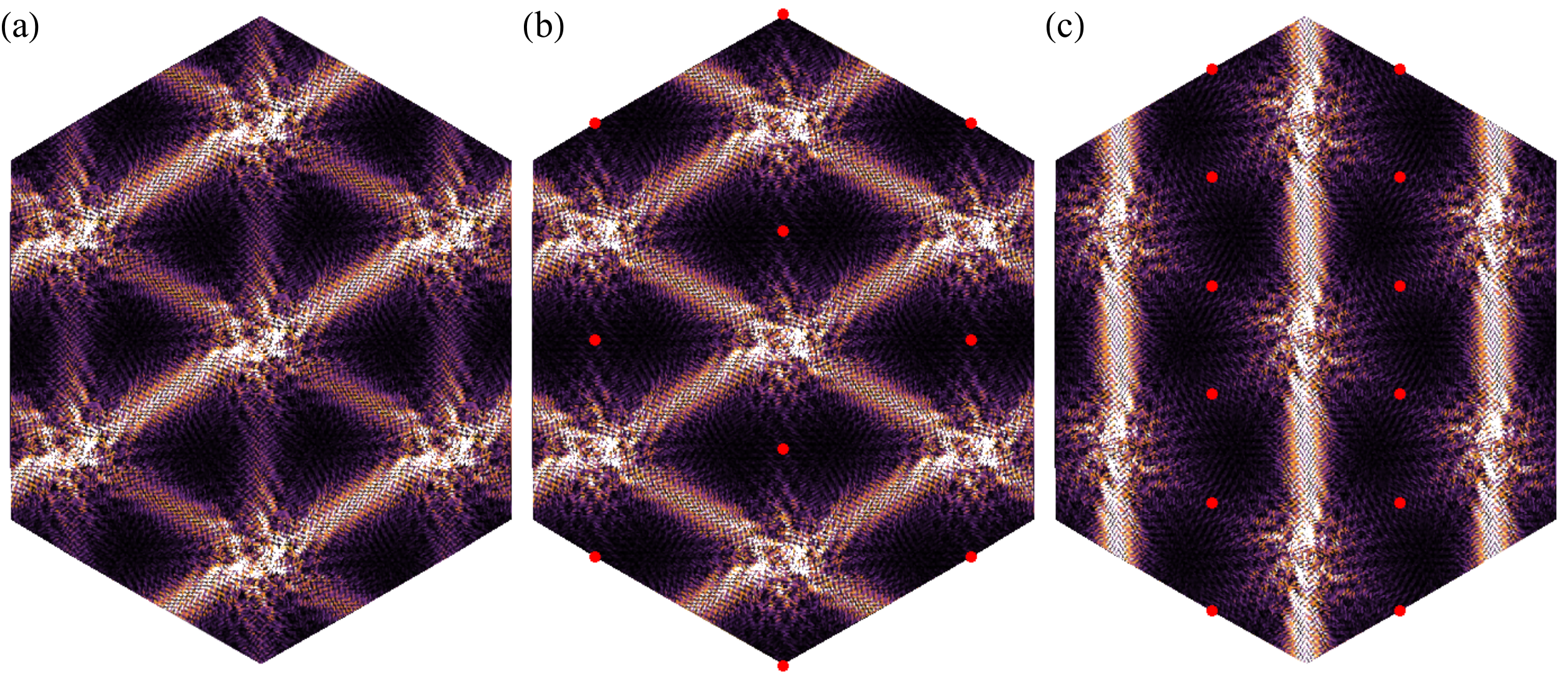} 
\caption{(Color online) (a) Spatial density of a helical network state similar to that of Fig. \ref{fig:eigenstates}(b), computed within a tight binding model of a biased twisted bilayer. (b,c) The same state in the presence of isolated atomic impurities adsorbed onto the top layer at positions marked by the red dots. Intervalley scattering on the defects suppresses the transmission through the corresponding link.
}
   \label{fig:circuit}
\end{figure}

The emergence of this extended network of helical states is very suggestive in relation to the possibility of fabricating circuits with topologically protected interconnects. Devices built this way are expected to be energetically more efficient than equivalent semiconductor-based systems, due to the fact that helicity prevents dissipation by momentum scattering \cite{Qi:RMP11}. A network of helical channels connecting logic gates must however be designed with high flexibility to achieve such goal. While, unlike edge channels in two-dimensional topological insulators, the present helical network spans the whole bilayer, it is not immediately clear whether its connectivity could be flexibly tailored at will into a specific circuit design. The key to this problem is that, as mentioned, the topological protection of helical channels is \emph{weak}, and can be broken by inducing local intervalley scattering. Certain defects, such as atomic vacancies or small adsorbates, are very efficient valley scatterers \cite{Matis:PRB13}. It turns out that a single hydrogen adsorbate can completely block a link in the helical network. We confirm this idea by tight-binding simulations on the full twisted bilayer lattice, see Fig. \ref{fig:circuit}. (Since the continuum model is valley-preserving by construction, it is not the best approach for such simulations.) All the technical details of the tight-binding simulation can be found in Refs. \onlinecite{Lofwander:PRB13, Prada:JOCE13}. For numerical efficiency, we introduced an identical set of atomic defects on each of the Moiré unit cells (red dots), and computed their effect on a helical network state similar to that of Fig. \ref{fig:eigenstates}(b). We found that any link with a single atomic defect becomes efficiently blocked within certain energy windows (sub-band crossings of from opposite valleys), which allows for a reconfiguration of the network into any desired two-dimensional connectivity pattern, as long as any undesired adsorbates are annealed away.


In conclusion, we have shown that twisted bilayer graphene under interlayer bias develops a novel spectral structure at low twist angles and realistic values of the bias, as a consequence of the depletion of AB' and BA'-type regions in the Moiré pattern. The periodic modulation of the local band topology gives rise to the formation of a delocalized helical network of protected states. These run along AB'/BA' boundaries, otherwise known as stacking solitons. Electrons propagate without dissipation along the links of the network, as long as disorder does not mix valleys. Conversely, special types of defects, such as hydrogen adsorbates or vacancies, can shut off transmission through a link due to valley scattering at precise energies, which leads to a gap opening in the link's one-dimensional helical dispersion. Thus, controlled adsorbate deposition could be used to modify the connectivity of the network at will, and create custom dissipationless topological circuits. 

In a remarkable experiment, Alden et al. \cite{Alden:PNAS13} recently showed that an array of stacking solitons does indeed form in graphene bilayers, with the largest typical periods approaching $1~\mathrm{\mu m}$, and are even mobile under the right conditions. Elastic deviations from the perfectly periodic model used here were observed, but being smooth on the scale of the lattice constant, the TPH states under bias are not destroyed by such distortions. Quite the contrary; as topological states, they become reinforced, since elastic deformations tend to maximise the size of the depleted AB' and BA' areas, and reduce overlaps. With a Moiré period distance of $\sim 100$nm, as in the experiment, we estimate a threshold interlayer bias of $\sim 45$meV. We thus believe that the unique Moiré superlattices of twisted bilayers are an ideal platform for studying the physics of electronic topological networks, such as the one considered in this work.

\acknowledgements
We gratefully acknowledge discussions with T. Stauber and L. Brey. We acknowledge financial support from the Spanish Ministry of Economy (MINECO) through the Ramón y Cajal programme, Grant no. FIS2010-21883 (E. P.), and Grant no. FIS2011-23713, and from the European Research Council Advanced Grant, contract 290846 (P. S-J.).

\bibliography{biblio}

\pagebreak

\appendix
 
\section{\large Appendix} 
  \begin{figure}[h!] 
\centering 
\includegraphics[width=\columnwidth]{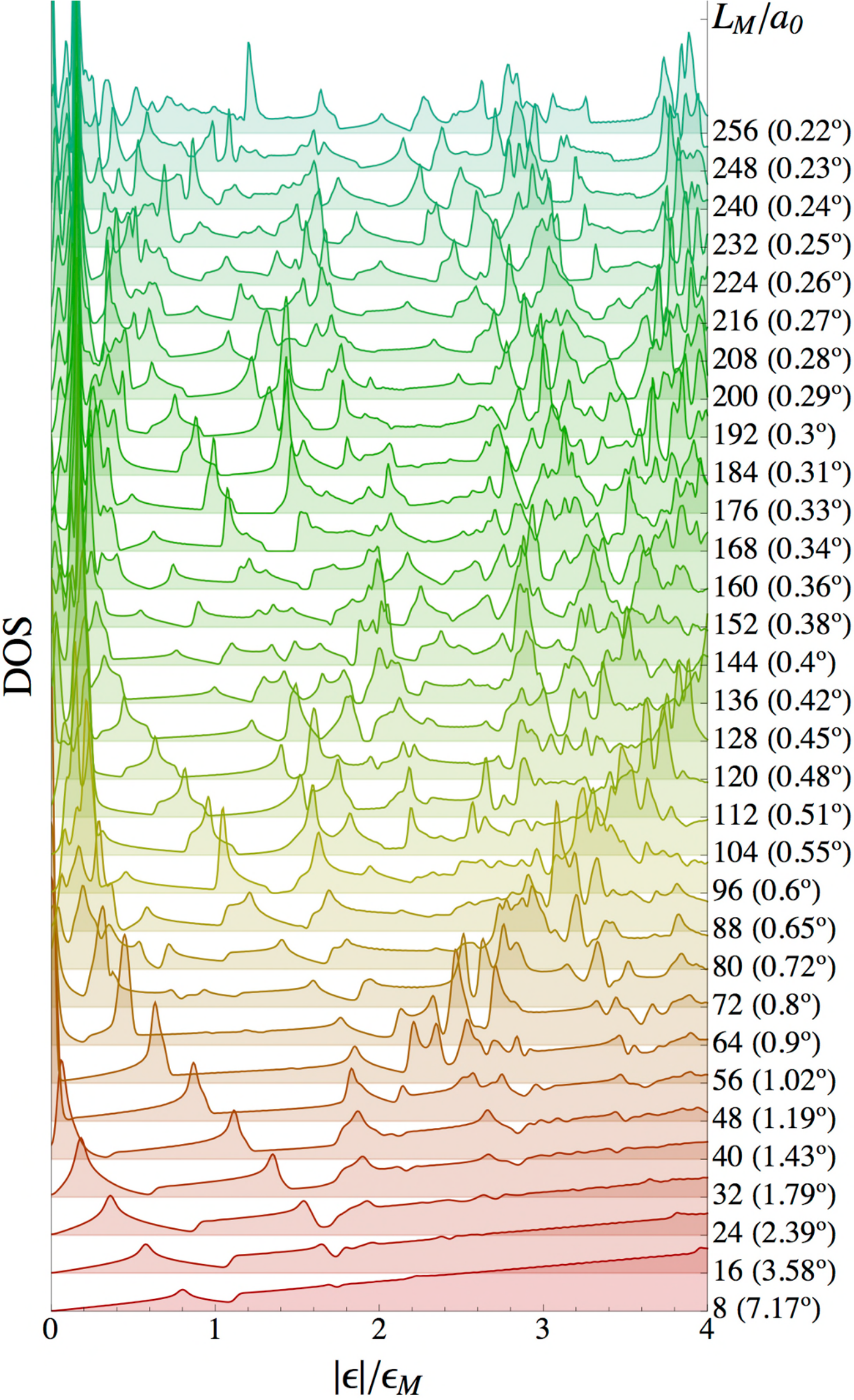} 
\caption{(Color online) Density of states of an unbiased twisted bilayer with decreasing twist angle $\theta$. $L_M$ is the Moiré period corresponding to each $\theta$, in parenthesis.
}
   \label{fig:DOS}
\end{figure}

 \begin{figure} 
\centering
\includegraphics[width=\columnwidth]{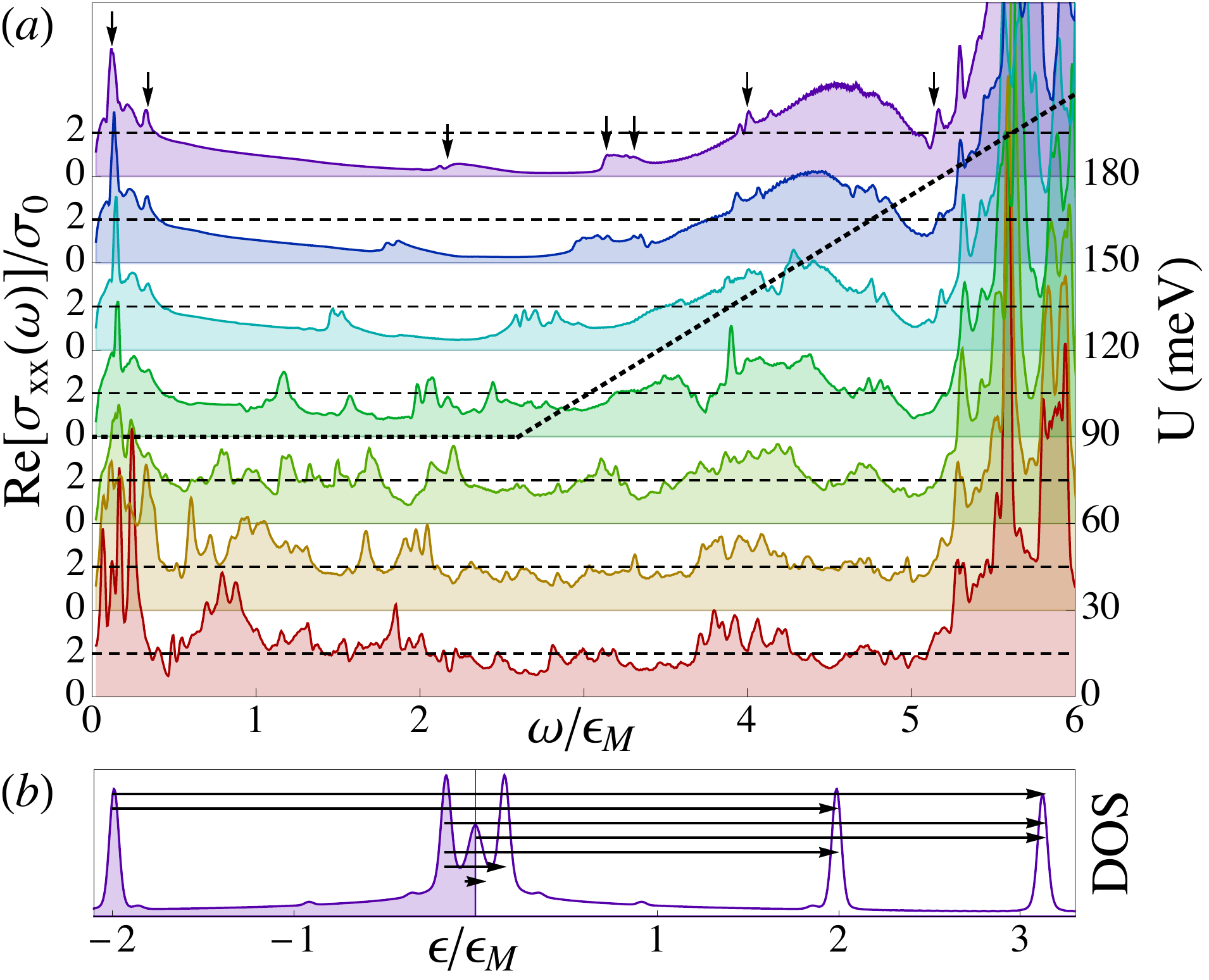} 
\caption{(Color online) (a) Zero-temperature optical conductivity $\mathrm{Re}[\sigma(\omega)]$ at neutrality ($\mu=0$) for increasing bias $U$ and twist angle like in Fig. 2 of the main text. It exhibits complex fluctuations around $2\sigma_0=\pi e^2/h$ at small $U$ (dashed line). As the system transitions into the helical network phase (above thick dotted line), the fluctuations are transformed into a smooth, non-monotonous background corresponding to transitions between helical states, with weak superimposed resonances from transitions between localised states. Some of the latter [arrows in (a)] are shown in (b) for $U=180$ meV and increasing frequency [arrows from bottom to top].
}
   \label{fig:sigma}
\end{figure}

The density of states (DOS) depends only on the density of energy eigenvalues of the system. Fig. \ref{fig:DOS} shows the evolution of the DOS in an unbiased twisted bilayer with the twist angle $\theta$. Other equilibrium observables, such as the optical conductivity $\sigma_{xx}(\omega)$, contain additional information related to  matrix elements between different eigenstates. It can be computed using the Kubo formula
\begin{equation}
\sigma_{xx}(\omega)=\frac{ie^2}{\omega}\int d\epsilon d\epsilon' \frac{n_F(\epsilon)-n_F(\epsilon')}{\hbar \omega-\epsilon'+\epsilon+i 0}F(\epsilon,\epsilon'),
\end{equation}
where $F(\epsilon,\epsilon')=\mathrm{Tr}\left[\delta(\epsilon-H)j_x\delta(\epsilon'-H)j_x\right]$, and $j_x$ is the current operator in the $x$ direction. Quite famously, the low-frequency limit of the optical conductivity is universal in a neutral graphene monolayer $\sigma_{xx}(\omega\to 0)=\sigma_0\equiv\frac{\pi}{2}\frac{e^2}{h}$, which corresponds to a universal infrared transmission at normal incidence $T=(1+\pi\alpha/2)^{-2}\approx 97.7\%$, in terms of the fine structure constant $\alpha$ \cite{Stauber:PRB08a,Nair:S08}.
In an unbiased, large-angle twisted bilayer, $\sigma_{xx}(\omega)$ exhibits features associated to the van-Hove singularities \cite{Moon:PRB13,Tabert:PRB13,Zou:PRL13}. As the system enters the low-angle regime, the optical conductivity exhibits a complexity similar to the DOS, in the form of fluctuations around a rough average of $2\sigma_0$ (which corresponds to two decoupled monolayers). Under an increasing interlayer bias, however, $\sigma_{xx}(\omega)$ becomes remarkably simple, like the DOS, see Fig. \ref{fig:sigma}. It evolves into a smooth background that is non-monotonous in $\omega$, and which falls below $2\sigma_0$, representing the optical absorption of the delocalised helical network. This background exhibits superimposed peaks that reflect transitions between sharp localised states [see arrows in panels (a) and (b)]. Such peaks, however, are much suppressed as compared to those in the DOS, due to the small overlap of localised eigenfunctions under the action of the current operator $j_x$. Hence, absorption due to the helical network becomes the dominant feature of $\sigma_{xx}(\omega)$ at low frequencies. For frequencies larger than $|U|/2$, complex features are still present, of a magnitude similar to those of the unbiased case.

\end{document}